\documentclass[showpacs,preprintnumbers,amsmath,amssymb,twocolumn]{revtex4-1}

\usepackage{graphicx}
\usepackage{dcolumn}
\usepackage{bm}

\newcommand{\be}{\begin{equation}}
\newcommand{\ee}{\end{equation}}
\newcommand{\lra}[1]{\langle #1 \rangle }

\begin{document}
\title{Reaction Spreading on Graphs}

\author{Raffaella Burioni}\affiliation{Dipartimento di Fisica and INFN, Universit\`a di Parma,Parco Area delle Scienze 7/A, 43100 Parma, Italy }
\author{Sergio Chibbaro}\affiliation{Institut D'Alembert University Pierre et Marie Curie, 4, place jussieu 75252 Paris Cedex 05}
\affiliation{CNRS UMR 7190, 4, place jussieu 75252 Paris Cedex 05}
\author{Davide Vergni}\affiliation{Istituto Applicazioni Calcolo, CNR, V.le Manzoni 30, 00185, Rome, Italy}
\author{Angelo Vulpiani}\affiliation{Dipartimento di Fisica, Universit\`a ``La Sapienza'' and ISC-CNR, Piazzale Aldo Moro 2, I-00185 Roma, Italy}

\begin{abstract} 
We study reaction-diffusion processes on graphs through an extension
of the standard reaction-diffusion equation starting from first
principles.  We focus on reaction spreading, i.e. on the time
evolution of the reaction product, $M(t)$.  At variance with pure
diffusive processes, characterized by the spectral dimension, $d_s$,
for reaction spreading the important quantity is found to be the
connectivity dimension, $d_l$. Numerical data, in agreement with
analytical estimates based on the features of $n$ independent random
walkers on the graph, show that $M(t) \sim t^{d_l}$. In the case of
Erd\"{o}s-Renyi random graphs, the reaction-product is characterized
by an exponential growth $M(t) \sim e^{\alpha t}$ with $\alpha$
proportional to $\ln \lra{k}$, where $\lra{k}$ is the average degree
of the graph.
\end{abstract}

\maketitle

A huge variety of different problems in chemistry, biology 
and physics deal with reactive species in non trivial 
substrates~\cite{Murray}.
Seminal works on reaction and diffusion dynamics date back to the
Fisher-Kolmogorov-Petrovskii-Piskunov (FKPP) model~\cite{FKPP}
\begin{equation}
   \partial_t \theta  =
   D \Delta \theta + f(\theta)\,,
   \label{eq:rd}
\end{equation}
where $D$ is the molecular diffusivity, $f(\theta)$ describes the
reaction process and the scalar field $\theta$ represents the
fractional concentration of the reaction products.
Afterward, reaction-transport dynamics attracted a considerable interest 
for their relevance in a large number of chemical,
biological and physical systems~\cite{Murray}.\\
Complex networks are a recent branch of graph
theory becoming very important for different disciplines ranging from
physics to social science, from biology to computer science
\cite{Barabasi1999}.
Although there exist an impressive amount of works on the study of
both complex networks and reaction-transport processes, as far as we
know, a general attempt to extend Eq.~(\ref{eq:rd}) on graphs 
and complex networks is still lacking.

There are two main approaches to study reaction dynamics on graphs.
One concerns agent based models (Lagrangian description) in which
random walkers move on the graph and interact, with a given reaction
rule, when they occupy the same site at the same
instant~\cite{Kopelman1984}.  A different approach is based on a
mesoscopic description of the reaction dynamics~(\ref{eq:rd}) in which
diffusion is modified introducing a proper transport term
taking into account the
feature of the media in which the dynamic takes place~\cite{Mendez2010}. 
A particular approach in the
mesoscopic description of the dynamics (used, e.g., in the recent
field of epidemic spreading~\cite{Vespignani2001}), is to use a mean-field approximation 
in which a renormalized reaction term takes into account the network characteristics.

The goal of this letter is to study the reaction spreading 
on graphs extending model~(\ref{eq:rd}).
In the presence of more general transport processes, the diffusion 
term $D\Delta \theta$ in Eq.~(\ref{eq:rd}), 
can be replaced by a suitable linear operator $\hat{L}$.
A general evolution equation for $\theta$ is:
\begin{equation}
   \partial_t \theta = 
      \hat{L} \theta + \frac{1}{\tau}f(\theta)\,\,.
   \label{eq:evol}
\end{equation}
where we write explicitly the typical time scale, $\tau$, of the
reaction process.
An important class of processes of this type is the
advection-reaction-diffusion (ARD), where $\hat{L}=-\mathbf{u} \cdot
\mathbf{\nabla}+D \Delta$. 
Another interesting case is ruled by the
effective diffusion operator $\hat{L}=\frac{1}{r^{d_{\tiny \mbox{f}}-1}}
\frac{\partial}{\partial r}\left(k(r)r^{d_{\tiny \mbox{f}}-1}
\frac{\partial}{\partial r}\right)$~\cite{Procaccia1985} 
suitable to study reaction dynamics on fractals~\cite{Mendez2010}.

Equation (\ref{eq:evol}) is constituted by two terms: 
the transport term, $\hat{L} \theta$, and
the non-linear local reaction, $f(\theta)/\tau$.  In the limit case without
reaction, the link between the solution $\theta({\bf x},t)$
and a suitable stochastic process is quite clear: for instance,
if $f(\theta)=0$ and $\hat{L}=-\mathbf{u} \cdot
\mathbf{\nabla}+D \Delta$, Eq.~(\ref{eq:evol}) is nothing but the
Fokker-Plank equation associated to the Langevin equation 
 ${{\mathrm d}{\bf x} / {\mathrm d}t} = 
   {\bf u} + \sqrt{2D} {\mbox{\boldmath{$\eta$}}}.
   \label{eq:langevin}$
In general, even in the presence of reactive terms
and for general $\hat L$, it is possible
to write $\theta({\bf x},t)$
in terms of  trajectories 
using the Freidlin formula~\cite{Freidlin}:
\be
\theta(\mathbf{x},t)=\left \langle{\theta(\mathbf{x},0)
     \exp{\left(\frac{1}{\tau} \int_0^t \frac{f(\theta(\mathbf{x}(s;t),s))}
   {\theta(\mathbf{x}(s;t),s)} ds \right)}} \right \rangle
\label{eq:Frei}
\ee
where the average is performed over all the trajectories 
${\bf x}(s;t)$ starting in ${\bf x}(0)$
and ending in ${\bf x}(t;t)= {\bf x}$.
The possibility to write the generalization of
(\ref{eq:Frei}) for a generic diffusive process
has been discussed in~\cite{acvv01}. Following this approach
we can determine the dynamical equation of reaction diffusion
on graphs.

As a diffusion process, we considered diffusion on an undirected,
unweighted and connected graph $G = (V, E)$, where $V$ is the set 
of vertices of the graph (we consider a finite number, $N$, of vertices)
and $E$ is the set of edges connecting the vertices. 
The graph  can be  represented by its adjacency matrix 
$A_{ij}$ given by~\cite{bollobas1998}:
\begin{equation}
A_{ij}=\left\{
\begin{array}{cl}
1 & {\rm if } \ (i,j) \in E \cr
0 & {\rm if } \ (i,j) \not\in E \cr
\end{array}
\right .
\label{defA}
\end{equation}

The discrete Laplacian of the graph $\Delta_{ij}$~\cite{bollobas1998,bc05}
is defined by:
$\Delta_{ij} = A_{ij} - k_i \delta_{ij}$
where $k_i=\sum_j A_{ij}$, the number of nearest neighbors of $i$, 
is the degree of vertex $i$. 
Once the rate of the jump process, $w$,
is introduced, the diffusion term can be written as
\begin{equation}
\frac{d \theta_i}{dt}=w \sum_j \Delta_{ij}\theta_j\,,
\label{eq:dgraph}
\end{equation}
where $\theta_i$ is the concentration at vertex $i$.
Our goal is to add to this equation a reaction term:
\begin{equation}
\frac{d \theta_i}{dt}=w \sum_j \Delta_{ij}\theta_j+\frac{1}{\tau}f(\theta_i)\,.
\label{eq:rdgraph}
\end{equation}
The discrete-time version of the diffusion equation~(\ref{eq:dgraph}) 
is nothing but
the random walk process described by the master equation
\begin{equation}
\theta_n(t + \Delta t) = \sum_j P_{j\to n}^{(\Delta t)} \theta_j(t)~,
\label{eq:evt}
\end{equation}
where jumps occur at time $\Delta t, 2 \Delta t, \ldots ,$ and
the probability for a walker being
at vertex $i$ to jump to the vertex $j$ are given in terms of 
the adjacency matrix
\begin{equation}
\begin{array}{lcr}
P_{i\to j}^{(\Delta t)}=wA_{ij}\Delta t \;\; &\text{if}& i \ne j \\
P_{i\to i}^{(\Delta t)}=1-k_i w \Delta t \;\; &\text{if}& i = j 
\end{array}
\label{eq:pij2}
\end{equation}
The discrete-time version of the reaction term can be defined as a non
zero function only at discrete time when $\delta$-form impulses occur:
\begin{equation}
  f(\theta,t) = \sum_{n=-\infty}^{\infty} g(\theta)
                                \delta(t - n\Delta t)\,\Delta t \,\,,
   \label{eq:reactmap}
\end{equation}
where $g(\theta)$ is a suitable function. Such a choice for $f(\theta)$
allows us for a rigorous treatment of the discretization of the reaction term.
With the above assumption, for Eq.~(\ref{eq:rdgraph}) we have
\begin{equation}
\theta_n(t+\Delta t)=G_{\Delta t} \left (
   \sum_j P^{(\Delta t)}_{j\to n} \theta_j(t) \right )\,,
\label{eq:discreterule}
\end{equation}
where $G_{\Delta t}(\theta)=\theta+g(\theta)\Delta t$ is the assigned
reaction map.  It is worth noting that Eq.~($\ref{eq:discreterule}$) 
can be seen as a numerical
method when the integration of (\ref{eq:rdgraph}) is performed in two
steps: diffusion and then reaction~\cite{acvv01}.
The shape of the reaction map $G_{\Delta t}(\theta)$ depends on the
underlying chemical model.  For auto-catalytic pulled reactions (the FKPP
class, where, e.g., $g(\theta)=\theta(1-\theta)/\tau$), characterized 
by an unstable fixed point in $\theta = 0$ and a
stable one in $\theta=1$ (the scalar field $\theta$ represents
the fractional concentration of the reaction products; $\theta=1$
indicates the inert material, $\theta=0$ the fresh one, and $0 <
\theta < 1$ means that fresh materials coexist with products)  one
can use $ G_{\Delta t}(\theta) = \theta + \theta(1-\theta) \Delta t
/\tau$. In the following we shall consider this type of reaction.

The most important topological features of a graph can be related to
the spectral dimension, $d_s$, and the
connectivity dimension, $d_l$, (also called chemical dimension).
The former is related to diffusion processes on graphs and
can be defined in terms of the return probability $P_{ii}$ at site $i$
for a random walker by $d_s=\lim_{t\rightarrow\infty}-2 \frac{\ln
P_{ii}(t)}{\ln t}$, or equivalently in terms of the density of
eigenvalues of the Laplacian operator \cite{bc05}. 
The connectivity dimension measures the average number
of vertices connected to a vertex in at most $l$ link, as
$\#(l) \sim l^{d_l}$.
For graphs embedded in an Euclidean space also the fractal dimension
$d_{\tiny \mbox{f}}$~\cite{ccv} should be considered, describing the
scaling of the number of vertices in a sphere of radius $r$ in the
Euclidean space, as $\#(r)\sim r^{d_{\tiny \mbox{f}}}$.  The
connectivity and fractal dimension can be different and they are
related via the mapping between the two distances $r$ and
$l$~\cite{Havlin1984}.

\begin{figure}[!h]
\begin{center}
\includegraphics[scale=0.3]{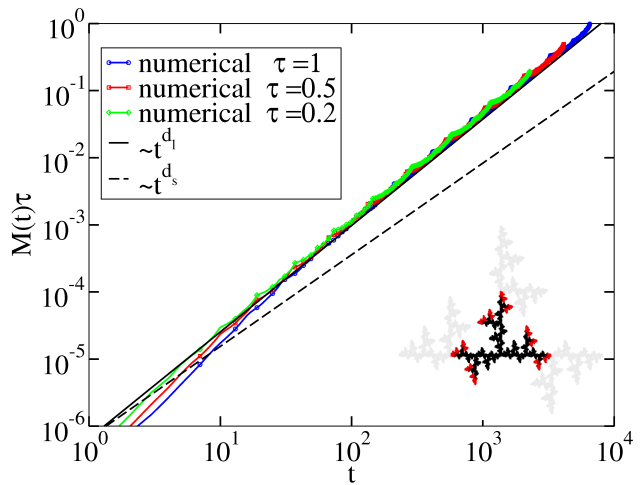}
\caption{The percentage of quantity of product times $\tau$,
$M(t)\tau$ vs $t$. 
Numerical results for Eq.~(\ref{eq:rdgraph}) with $w=0.5$
are compared to prediction $t^{d_l}$. For this graph 
$d_l=\ln 3/\ln 2\simeq 1.585$,
$d_l=2\ln 3/\ln 5\simeq 1.365$. Inset: Spreading on a T-fractal.}
\vspace{-0.5truecm}
\label{fig:1}
\end{center}
\end{figure}

As a typical example of undirected, unweighted and connected graph, 
we show in Fig.~\ref{fig:1} the reaction spreading
in the T-graph \cite{Hav_84}. 
The field $\theta$
is initialized to zero in each vertex except the central one
in which $\theta_i(0)=1$. Using Eq.~(\ref{eq:rdgraph}) we study the time evolution
of the system. An interesting observable to characterize the
spreading of the reaction is the percentage of 
the total quantity of the reaction product, i.e.,
$M(t)=\frac{1}{N}\sum_{i\in V}\theta_i(t)$
where $N$ is the total number of vertices. 
As clearly shown in Fig.~\ref{fig:1}, $M(t)$ grows as a power law
that can be interpreted as follows.
Starting from a single vertex with
$\theta_i(0)=1$, after $t$ step the number of vertices reached 
by the field is $\#(t)\sim t^{d_l}$. Therefore, in the limit
of very fast reaction, when each vertex reached by the field
is immediately burnt (i.e, $\theta_j \to 1$), we can expect:
\begin{equation}
M(t)\sim t^{d_l}.
\label{eq:dymobs}
\end{equation}
Fig.~\ref{fig:1} confirms that the connectivity dimension is the
relevant quantity for the reaction spreading on graph.  This behavior
can be also understood thinking of the asymptotic behavior of the
reaction process as determined by the spreading of the front in the
topological metric of the graph. In this case the characteristic time 
of reaction, $\tau$, appears only in the prefactor of the exponential.

Moreover, a theoretical argument further confirms the importance
of $d_l$. The analysis is based on an
analogy between reaction spreading and short time regime of the number of
distinct sites visited by $n$ independent random walkers after $t$
steps on a graph, $S_n(t)$~\cite{weiss}. This quantity can be computed as 
$S_n(t) = \sum_{j=0}^N 1-C_{0j}(t)^n$,
where $C_{0j}(t)$ is the probability that a walker starting from site
$0$ has not visited site $j$ at time $t$, 
the sum is over all the $N$ sites of the graph
dropping the dependence on the starting site $0$. 
When the number of walkers is large ($n\to\infty$),
$C_{0j}(t)^n$ tends to zero if site $j$ has a non zero probability 
of being reached in $t$ steps. In this limit, $S_n(t)$ represents all the
sites which have nonzero probability of being visited by
step t and, as $t$ is equal to the connectivity distance, 
$S_n(t) \sim t^{d_l}.$
This is precisely the regime observed in the reaction spreading
(see Eq.~(\ref{eq:dymobs}) and Fig.~\ref{fig:1}). 
An estimate of the validity of the short time regime is given in terms
of the smallest non zero occupation probability on the graph at time
$t$, $P_m=\lra{k}^{-t}$, being $\lra{k}$ the average degree of the
graph, i.e., the mean number of link for each vertex. As the short
time regime is supposed to hold as long as $nP_m \gg 1$, one obtains
that the reaction spreading regime is observed up to times $\bar t \sim
\ln n$.  On the other hand, the asymptotic regime is dominated by the
number of distinct visited sites by a walker, that is $S_n(t) \sim
t^{d_s/2} $ in graphs with compact exploration $d_s<2$, or simply by
$t$ on graphs with $d_s >2$ \cite{weiss} . In the case of very fast
reaction regime, the front can be considered as equivalent to an
infinite number of walkers, hence the asymptotic regime is
never reached, leaving the dynamics to be governed by the sole $d_l$.

As for the spectral dimension, it is the relevant quantity when dealing
with random-walk dynamics~\cite{Kopelman1984} and in some reaction
diffusion processes. For instance in~\cite{bettolo}
Eq.~(\ref{eq:rdgraph}) has been studied for coarsening processes where
$f(\theta)$, at variance with our case, has a bistable structure.
Moreover a further argument confirms the minor role of the spectral
dimension in the case of reaction-diffusion dynamics using FKPP
reaction terms.  When dealing with standard diffusion ($\langle
x^2(t)\rangle \sim t$) it is possible to show~\cite{acvv01} that the
spreading dynamics is the same displayed by the standard
reaction/diffusion problem~(\ref{eq:rd}), i.e., $M(t)\sim t^d$ (where
$d$ is the dimension of the space).  On the other hand, the presence
of anomalous diffusion ($\langle x^2(t)\rangle \sim t^{2\nu}$ with
$\nu \neq 1/2$) does not implies that the spreading is anomalous: 
case exists~\cite{acvv01} in which diffusion is anomalous 
but reaction spreading is standard.

The same behavior displayed in Figure~\ref{fig:1} has been observed
in several other self-similar graphs (e.g., Vicsek and Sierpinski carpet,
not shown here), confirming the leading role of $d_l$.  In the case
of percolation clusters, the importance of the connectivity dimension, 
and the difference between
connectivity dimension and fractal dimension was previously
shown~\cite{Havlin1984}. 
\begin{figure}[!h]
\begin{center}
\includegraphics[scale=0.3]{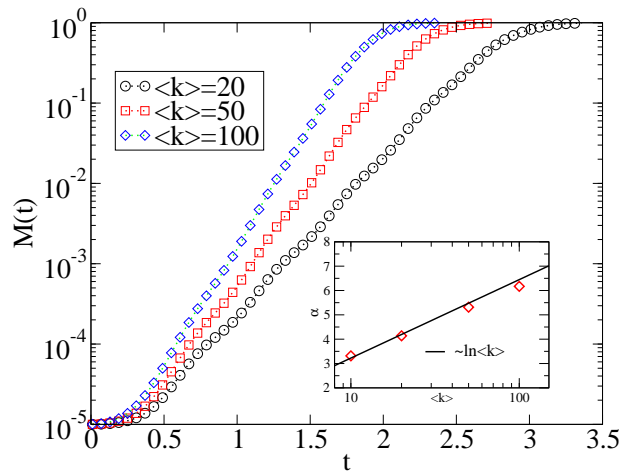}
\caption{Reaction product $M(t)$ vs $t$ for Erd\"{o}s-Renyi graphs.
Three results for different average degree of
connectivity are shown in the fast reaction regime ($\tau=0.1$). 
The reaction spreading follows an exponential behavior $M(t)\sim t^\alpha$,
where $\alpha$ depends on $\lra{k}$ as shown in the inset.}
\label{fig:2}
\end{center}
\end{figure}

Now we focus on the behavior of Eq.~(\ref{eq:rdgraph}) for
Erd\"{o}s-Renyi (ER) random graphs~\cite{bollobas1998} for which
$d_l=\infty$.  In the ER graphs two vertices are connected with
probability $p$.  We choose $p>\frac{\ln(N)}{N}$ so that the graph
contains a global connected component. The average degree of the graph
is $\lra{k}=p(N-1)\,.$
On ER graphs the number of points in a sphere
of radius $l$ grows exponentially, $\#(t) \sim e^{c\,t}$, hence
we expect a similar behavior
for the spreading process: 
\be
M(t)\sim e^{\alpha t},
\ee
as shown in Fig~\ref{fig:2}.  
If $\lra{k}$ is large and the reaction is slow enough
we have a two steps mechanism: first there is a rapid diffusion on the
whole graph, then the reaction induces an increase of $\theta_i$. This
leads to a simple mean field reaction dynamics, 
$\partial_t\rho(t)=\rho(t)(1-\rho(t))/\tau$, where
$\rho$ is the average value of $\theta_i$ on the graph. In this case
$\alpha=1/\tau$ as clearly observed in numerical simulations
(not shown here). 

In the much more interesting case of fast reaction, at
each time step the number of sites invaded is proportional to the
average degree of the graph, so that after $t$ steps we have:
\be 
M(t)\sim (C_1\lra{k})^t=e^{C_2\ln\lra{k}t}\,,
\ee
leading to $\alpha \sim \ln\lra{k}$, see inset of Fig~\ref{fig:2}.

\begin{figure}[!h]
\begin{center}
\includegraphics[scale=0.3]{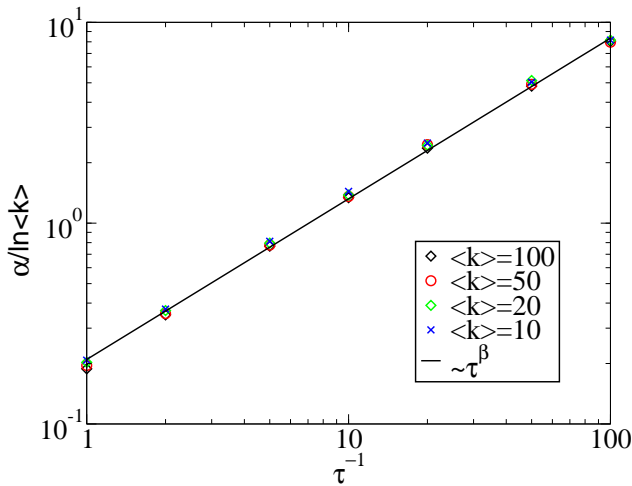}\\
\caption{The scaling exponent $\alpha$ 
normalized with $\ln\lra{k}$, as a
function of the inverse of reaction-time $\frac{1}{\tau}$. 
The straight line indicate $\tau^\beta$ with $\beta=-0.8$.}
\label{fig:3}
\end{center}
\end{figure}

Furthermore, at variance with the case of graphs with finite $d_l$, at
least in the case of fast reaction and FKPP reaction term, $\tau$
plays an important role since $C_2$ is a function of $\tau$.
Fig.~\ref{fig:3} shows the dependence of the exponential behavior of
reaction spreading rescaled with $\ln\lra{k}$ as a function of $\tau$.
We can fit the dependence of the curve on $\tau$ with
$\alpha(k,\tau)\simeq C\tau^\beta\ln\lra{k}$,
with $\beta\simeq -0.8$.
This scaling can be related to a mean field-like equation of the type:
\be
\partial_t\rho(t)=C\tau^\beta\ln(\lra{k})\rho(t)(1-\rho(t))\,.
\ee

We have considered a general model for reaction-diffusion dynamics on
graphs, allowing for a general and detailed treatment of the diffusive
and reaction terms. We study only large systems in which the
asymptotic scaling for the reaction spreading is well defined.  On the
other hand, although the spreading dynamics on small systems is
certainly a very interesting issue, it deserves careful attention and
it is beyond the scope of the present work.  In fact, even in the
absence of reaction (i.e., pure diffusion) in small systems the
boundaries can induce rather complicated behaviours~\cite{marchesoni}.\\
On undirected and finite dimensional graphs, we
found that a major role in the reaction spreading is played by the
connectivity dimension, which rules the asymptotic of the reaction
product as a function of time.  On random graphs with infinite
connectivity dimension, the reaction spreading shows an exponential
behavior, whose scaling depends on the average degree of the graph.
In this case, we obtain two mean-field like equations, one in the slow
reaction limit and one in the fast reaction limit. In particular, in
the fast reaction case, non-trivial dependence on both the average
degree of the graph and the reaction characteristic time is shown.
Our approach could be therefore suitable for a rigorous derivation of
mean field like equations in more complex topologies.

\end{document}